\documentclass[prd,twocolumn,showpacs,floatfix,amsmath,nofootinbib,amssymb,floatfix]{revtex4}
\usepackage{graphicx,color,dcolumn,booktabs,bm}
\usepackage{longtable,lscape}
\usepackage{txfonts}
\usepackage{overpic}
\usepackage{amssymb}
\usepackage{indentfirst}
\usepackage{feynmf}   
\usepackage{slashed}  
\usepackage{cases}
\usepackage{color}
\usepackage{multirow}
\usepackage{epstopdf}
\usepackage{enumerate}
\usepackage{graphicx,color,dcolumn,booktabs,bm}
\usepackage[colorlinks, citecolor=blue,anchorcolor=red,menucolor=red, linkcolor=red,filecolor=red,urlcolor=blue,frenchlinks=red]{hyperref}

\graphicspath{{Figures/}} %

\begin{document}

\title{Investigation of $\Lambda_Q$ and $\Xi_Q$ baryons in the heavy quark-light diquark picture}
\author{Bing Chen$^1$}\email{theophys@163.com}
\author{Ke-Wei Wei$^1$\footnote{Corresponding author}}\email{weikw@hotmail.com}
\author{Ailin Zhang$^2$}\email{zhangal@staff.shu.edu.cn}
\affiliation{$^1$Department of Physics, Anyang Normal University,
Anyang 455000, China\\
$^2$Department of Physics, Shanghai University, Shanghai 200444,
China}


\begin{abstract}
We apply a new mass formula which is derived analytically in the
relativistic flux tube model to the mass spectra of $\Lambda_Q$ and
$\Xi_Q$ (\emph{Q} = \emph{c} or \emph{b} quark) baryons. To this
end, the heavy quark-light diquark picture is employed. We find that
all masses of the available $\Lambda_Q$ and $\Xi_Q$ states can be
understood well. The assignments to these states do not appear to
contradict the strong decay properties. $\Lambda_c(2760)^+$ and
$\Xi_c(2980)$ are assigned to the first radial excitations with $J^P
= 1/2^+$. $\Lambda_c(2940)^+$ and $\Xi_c(3123)$ might be the
2\emph{P} states. The $\Lambda_c(2880)^+$ and $\Xi_c(3080)$ are the
good 1\emph{D} candidates with $J^P = 5/2^+$. $\Xi_c(3055)$ is
likely to be a 1\emph{D} state with $J^P = 3/2^+$.
$\Lambda_b(5912)^0$ and $\Lambda_b(5920)^0$ favor the 1\emph{P}
assignments with $J^P = 1/2^-$ and $3/2^-$, respectively. We propose
a search for the $\tilde{\Lambda}_{c2}(5/2^-)$ state which can help
to distinguish the diquark and three-body schemes.

\end{abstract}
 \maketitle

\section{Introduction}

The so-called missing resonances problem has never been understood
for the baryon physics. In the constituent quark model, a baryon
contains three confined quarks. If this is true, the predicted
states are much more than observations. For the light flavor
baryons, Galat$\grave{a}$ and Santopinto have pointed out that the
established and tentative states listed by the Particle Data Group
(PDG)~\cite{PDG} are much less than the theoretical
predictions~\cite{Sant1}. For the heavy baryons, we take the charmed
baryon as an example. More than 50 states are allowed for the
$\Lambda_c$ and $\Sigma_c$ baryons up to $N = 2$ shell in the
three-body picture. But only 9 $\Lambda_c$ and $\Sigma_c$ candidates
have been listed by PDG at present~\cite{PDG}.

A heuristic and possible solution to this problem is to introduce
``diquarks''~\cite{Klempt,Crede}. Since the degree of freedom of the
two quarks in diquark is frozen, the number of excited states shall
be greatly reduced. For the nonstrange light baryons, the good
descriptions of the masses up to 2 GeV have been provided by
different quark-diquark models~\cite{Sant2,Sant3,Gutierreza}. Based
on a QCD motivated quark potential model~\cite{Ebert1}, the mass
spectra of heavy baryons have been calculated in the heavy
quark-light diquark picture. There the light diquarks were treated
as completely relativistic and the heavy quarks were expanded in
$v/c$ up to the second order. An improved method without any
expansions was adopted later~\cite{Ebert2}. Masses for the higher
excited heavy baryon states were presented. Based on the
predictions, the Regge trajectories for orbital and radial
excitations were constructed. The linearity, parallelism, and
equidistance were verified. Most importantly, the authors concluded
that ``all available experimental data of heavy baryons fit nicely
to the constructed Regge trajectories''~\cite{Ebert2}. The strong
decays have not been discussed in Refs.~\cite{Ebert1,Ebert2}.

To our knowledge, only Refs.~\cite{Ebert1,Ebert2} focused on the
mass spectra of the high excited heavy baryons systematically in the
quark-diquark picture. In these two works, a relativistic quark
potential model was used. So it is required to test the
quark-diquark picture for the heavy baryons by other models.

The relativistic flux tube (RFT) model is not a potential approach
because the interaction is mediated by a dynamical
tube~\cite{Olsson1,Olsson2}. Selem and Wilczek employed the
relativistic flux tube model to investigate whether there are
diquark in the baryons~\cite{Selem}. For the heavy baryons, the
following mass formula
\begin{equation}\label{eq1}
E = M+\sqrt{\frac{\sigma L}{2}}+2^{1/4}\kappa L^{-1/4}\mu^{3/2}.
\end{equation}
was obtained by the computer simulations~\cite{Selem}. Here
\emph{E}, \emph{M}, and $\mu$ refer to the masses of baryon system,
heavy quark, and light-diquark respectively. \emph{L} is the orbital
angular momentum. The string tension is denoted by $\sigma/(2\pi)$.
The parameter $\kappa$ is dependent on $\sigma$:
$\kappa\equiv\frac{2\pi^{1/2}}{3\sigma^{1/4}}$. Obviously, the
formula above does not include the ground state because of the
singularity. In addition, the spin-dependent interactions have not
been incorporated. This formula has been used to study the mass
spectrum of $D$, $D_s$~\cite{zhang1,zhang2} and
$\Lambda_c^+$~\cite{zhang3}, where the spin-orbit interactions have
been taken into account.

In this work, a different mass formula for the heavy-light hadrons
will be analytically derived within the relativistic flux tube
model. Then we will apply the new formula to the $\Lambda_Q$ and
$\Xi_Q$ baryons, where the two light quarks are treated as a scalar
diquark. The spin-orbit interaction will be borrowed from the
QCD-motivated constituent quark models. We will also discuss the
decays of 1\emph{D} candidates for completeness.

The paper is organized as follows. In Section \ref{sec2}, we derive
the mass formula of the heavy-light hadrons in the RFT model. In
Section \ref{sec3}, the spectrum of $\Lambda_Q$ and $\Xi_Q$ baryons
are discussed. In Section \ref{sec4}, we further explore how to test
the diquark and three-body schemes by experiments in future. The
last section contains the summary and outlook.

\section{The mass formula of heavy-light hadrons in the RFT model}\label{sec2}
In the RFT model, the confined quarks in a hadron are assumed to be
connected by the relativistic color flux tube which carries both
energy and momentum~\cite{Olsson1}. The prototype of this model is
the Nambu-Goto QCD string model \cite{Nambu,Susskind,Goto}. The RFT
model has been studied carefully by Olsson \emph{et
al}.~\cite{Olsson3,Olsson4,Olsson5,Olsson6,Olsson7,Olsson8}. An
interesting research topic of this model is to reproduce the Regge
trajectories behavior of different
hadrons~\cite{Selem,Olsson3,Olsson4}. Besides the heavy-light
hadrons, the RFT model was also applied to the
charmonium~\cite{Burns}, pentaquark~\cite{penta}, and
glueball~\cite{glue}.

In the RFT model, the energy $\varepsilon$ and the angular momentum
\emph{L} of a hadron system are given as follows~\cite{Olsson1}
\begin{equation}\label{eq2}
\varepsilon = \sum_{i=1}^2[\frac{m_i}{\sqrt{1-(\omega
r_i)^2}}+\frac{T}{\omega}\int^{\omega
r_i}_0\frac{du}{\sqrt{1-u^2}}],
\end{equation}
and
\begin{equation}\label{eq3}
L = \sum_{i=1}^2[\frac{m_i \omega r_i^2}{\sqrt{1-(\omega
r_i)^2}}+\frac{T}{\omega^2}\int^{\omega
r_i}_0\frac{u^2du}{\sqrt{1-u^2}}].
\end{equation}

Here, we have omitted the velocity of light, \emph{c}, for
simplicity. The equations above have been derived rigorously from
the Wilson area law in QCD when the spin-dependent terms were
neglected~\cite{Brambilla}. Then the $m_i$ can be regarded as the
``current quark masses''. When we define the $v_i = \omega r_i$, the
Eqs. \ref{eq2} and \ref{eq3} become
\begin{equation}\label{eq4}
\varepsilon =
\sum_{i=1}^2(\frac{m_i}{\sqrt{1-v_i^2}}+\frac{T}{\omega}\arcsin
v_i),
\end{equation}
and
\begin{equation}\label{eq5}
L =
\sum_{i=1}^2[\frac{m_i}{\omega}\frac{v_i^2}{\sqrt{1-v_i^2}}+\frac{T}{2\omega^2}(\arcsin
v_i-v_i\sqrt{1-v_i^2})].
\end{equation}

We assume $m_1 \ll m_2$ for heavy-light hadron systems, and define
$m_l = m_1/\sqrt{1-v_1^2}$, $m_Q = m_2/\sqrt{1-v_2^2}$. Since $m_l$
and $m_Q$ have included the relativistic effect, they might be
treated as the constituent quark masses.

As shown in the Section \ref{sec3}, the velocity of heavy quark in
the heavy baryons is about 0.5\emph{c}. But the light diquark is
ultrarelativistic. We take the rest of $v_1$ as 1 for approximation,
and expand Eqs. (\ref{eq4}) and (\ref{eq5}) up to the second order
in the parameter $v_2$. Then we obtain
\begin{equation}\label{eq6}
\varepsilon = m_Q + m_l + m_Qv_2^2+\frac{\pi T}{2\omega},
\end{equation}
and
\begin{equation}\label{eq7}
L = \frac{1}{\omega}(m_l + m_Qv_2^2+\frac{\pi T}{4\omega}).
\end{equation}
We have used the following relationship
\begin{equation}\label{eq8}
\frac{T}{\omega} = \frac{m_2 v_2}{1-v^2_2} \simeq m_Q v_2.
\end{equation}

Based on the Eqs.~(\ref{eq6}) and (\ref{eq7}), the spin-averaged
mass formula for the orbital excited states is obtained directly as
\begin{equation}\label{eq9}
(\varepsilon-m_Q)^2 = \frac{1}{2}\sigma L + (m_l + \zeta_Q)^2,
\end{equation}
where $\sigma = 2\pi T$ and $\zeta_Q = m_Qv_2^2$. Eq.~(\ref{eq9})
above is the Chew-Frautschi formula for heavy-light systems. A
formula without the intercept was obtained in Ref.~\cite{Olsson1}.
Another formula with a different intercept was found as:
$(\varepsilon - m_2)^2 = \sigma L/2 + \sigma/6$~\cite{Baker}, where
the effect of string fluctuations was considered. The formulae in
Refs.~\cite{Olsson1,Baker} were achieved in the physical limits that
$m_1 \rightarrow 0$ and $m_2 \rightarrow \infty$. So it seems
unreasonable to apply them to the ordinary heavy-light hadrons for
the finite masses of quarks. The intercept of Eq.~(\ref{eq9}) is
also different from the Eq.~(\ref{eq1}) because the singularity no
longer appears. Thus, we expect that the Eq.~(\ref{eq9}) also
includes the case of $L = 0$. Of course, the angular momentum $L$
can not be understood in the classical picture when we use
Eq.~(\ref{eq9}) to describe the heavy-light hadrons.

Since the confined quarks are treated as spinless particles in the
RFT model, the spin-dependent terms can not be given by the
semi-classical way above. For $\Lambda_Q$ and $\Xi_Q$, the
spin-dependent interactions are much simple because the primary
couplings exist between the orbital angular momentum and the spin of
heavy quark. With the axial-vector diquark (see Section \ref{sec3}),
$\Sigma_Q$ and $\Xi'_Q$ have more complicated hyperfine structure.
For simplicity, we will only study the $\Lambda_Q$ and $\Xi_Q$
baryons in this work. The $\textbf{S}_Q \cdot \textbf{L}$ couplings
can be borrowed from the QCD-motivated quark potential models.
Similar to heavy-light mesons~\cite{Cahn}, the spin-orbit couplings
for $\Lambda_Q$ and $\Xi_Q$ have the form
\begin{equation}\label{eq10}
H_{so} =
\frac{4}{3}\frac{\alpha_s}{r^3}\frac{1}{m_dm_Q}~\textbf{S}_Q\cdot
\textbf{L}.
\end{equation}
Here the second and higher orders of $1/m_Q$ are ignored. $m_d$
refers to the mass of light diquark. $\alpha_s$ is the coupling
constant. Combing the Eqs.~(\ref{eq7}) and~(\ref{eq9}), the angular
velocity $\omega$ could be expressed as
\begin{equation}\label{eq11}
\omega = \frac{\varepsilon -\varepsilon_0}{2L} =
\frac{\sigma}{8\omega L}.
\end{equation}
Here, $\varepsilon_0$ denotes the energy of ground state $(L~=~0)$.
When the $r\omega = v_1+v_2$ is considered, we obtain
\begin{equation}\label{eq12}
\frac{1}{r} = \frac{\omega}{v_1 + v_2} = \frac{1}{v_1 +
v_2}\sqrt{\frac{\sigma}{8L}}.
\end{equation}
The orbital angular momentum of the heavy-light meson was shown as
$L \rightarrow \sigma r^2/8$ under the assumptions that the light
quark is ultrarelativistic, and the mass of heavy antiquark is
infinite~\cite{Olsson5}. In other words, the limits of $v_1 = 1$ and
$v_2 = 0$ were used in Ref~\cite{Olsson5}. By substituting the
expressions~(\ref{eq12}) into Eq.~(\ref{eq10}), we find
\begin{equation}\label{eq13}
H_{so} = \frac{1}{3\times2^{5/2}}\frac{\alpha_s}{(v_1 +
v_2)^3}(\frac{\sigma}{L})^{3/2}\frac{1}{m_dm_Q}\hat{s}_Q\cdot
\hat{L}.
\end{equation}

Based on the numerical analysis~\cite{Olsson3,Olsson6}, and the
semi-classical quantization scheme~\cite{Olsson8}, the RFT model
implied that the linearity, parallelism, and equidistance may exist
in the Regge trajectories of the heavy-light hadrons. Inspired by
these results, we would like to extend Eq.~(\ref{eq9}) to the radial
excited heavy baryons
\begin{equation}\label{eq14}
(\varepsilon-m_Q)^2 = \frac{1}{2}\sigma (\lambda n + L) + (m_d +
\zeta_Q)^2.
\end{equation}
The coefficient, $\lambda$, will be determined directly by the
experimental data. Accordingly, the $\textbf{S}_Q \cdot \textbf{L}$
couplings term are revised as
\begin{equation}\label{eq15}
H^{so}_{nL} = \frac{1}{3\times2^{5/2}}\frac{\alpha_s}{(v_1 +
v_2)^3}(\frac{\sigma}{\lambda n +
L})^{3/2}\frac{1}{m_dm_Q}\hat{s}_Q\cdot \hat{L}.
\end{equation}

\section{Application}\label{sec3}
In this Section, the phenomenological Eqs.~(\ref{eq14})
and~(\ref{eq15}) will be applied for the $\Lambda_Q$ and $\Xi_Q$
baryons. For most of these baryons, the quantum numbers have not yet
been determined experimentally. Thus, they were usually prescribed
following the quark model predictions. Their different aspects have
been reviewed in detail in Refs.~\cite{Klempt,Crede}. Here, we
collect all $\Lambda_Q$ and $\Xi_Q$ candidates listed by
PDG~\cite{PDG} in Table \ref{tableI}. One notice that the mass gaps
of the corresponding $\Xi_c$ and $\Lambda_c^+$ are about
180$\sim$200 MeV. The mass gap of $\Xi_b (5790)$ and
$\Lambda_b(5620)$ is about 170 MeV. For illustrating such law
clearly, we present the $\Lambda_Q$ and $\Xi_Q$ baryons alongside in
Table \ref{tableI}. The phenomena of mass gaps reflects the similar
dynamics of the $\Lambda_Q$ and $\Xi_Q$ baryons (see Section
\ref{sec4}). In the following, we will discuss $\Lambda_Q$ and
$\Xi_Q$ in parallel.

\begin{table*}
\caption{The experimental information of the $\Lambda_Q$ and $\Xi_Q$
baryons. The average values of mass and decay width (in units of
MeV) are taken from PDG~\cite{PDG}. Here the predicted
$\Lambda_c(2860)^+$, $\Xi_b(6095)^-$, and $\Xi_b(6105)^-$ are listed
for comparison and completeness. The mass differences between
$\Lambda_Q$ and $\Xi_Q$ baryons are listed in the last column.}
\label{tableI}
\begin{tabular*}{\textwidth}{@{\extracolsep{\fill}}lllllllll@{}}
\hline
Names\hspace {0.4cm}     &\hspace {0.08cm}Status  & Mass                  & Width                &  Names\hspace {0.4cm} &\hspace {0.02cm}Status   & Mass                & Width    & \hspace {0.08cm}$\Delta$M~(MeV)   \\
\hline
$\Lambda_c(2286)^+$      &$\ast\ast\ast\ast$   &  $2286.46\pm0.14$     &  $-$                 & $\Xi_c(2468)^0$       &$\ast\ast\ast$  &  $2470.88^{+0.34}_{-0.80}$&  $-$                 & $184.42^{+0.37}_{-0.81}$ \\
$\Lambda_c(2595)^+$      &$\ast\ast\ast$       &  $2592.25\pm0.28$     &  $2.6\pm0.6$         & $\Xi_c(2790)^0$       &$\ast\ast\ast$  &  $2791.8\pm3.3$           &  $<12$               & $199.6\pm3.3$\\
$\Lambda_c(2625)^+$      &$\ast\ast\ast$       &  $2628.11\pm0.19$     &  $<0.97$             & $\Xi_c(2815)^0$       &$\ast\ast\ast$  &  $2819.6\pm1.2$           &  $<6.5$              & $191.5\pm1.2$ \\
$\Lambda_c(2765)^+$      &$\ast$               &  $2766.6\pm2.4$       &  $50$                & $\Xi_c(2980)^0$       &$\ast\ast\ast$  &  $2968.0\pm2.6$           &  $20\pm7$            & $201.4\pm3.5$\\
$\Lambda_c(2860)^+$      &$\cdots$             &  $\cdots$             &  $\cdots$            & $\Xi_c(3055)^+$       &$\ast\ast$      &  $3054.2\pm1.3$           &  $17\pm13$           & $\cdots$  \\
$\Lambda_c(2880)^+$      &$\ast\ast\ast$       &  $2881.53\pm0.35$     &  $5.8\pm1.1$         & $\Xi_c(3080)^0$       &$\ast\ast\ast$  &  $3079.9\pm1.4$           &  $5.6\pm2.2$         & $198.4\pm1.4$  \\
$\Lambda_c(2940)^+$      &$\ast\ast\ast$       &  $2939.3^{+1.4}_{-1.5}$& $17^{+8}_{-6}$      & $\Xi_c(3123)^+$       &$\ast$          &  $3122.9\pm1.3$           &  $4\pm4$             & $183.6^{+1.9}_{-2.0}$ \\
$\Lambda_b(5619)^0$      &$\ast\ast\ast$       &  $5619.4\pm0.6$       &  $-$                 & $\Xi_b(5790)^-$       &$\ast\ast\ast$  &  $5791.1\pm2.2$           &  $-$                 & $171.7\pm2.3$ \\
$\Lambda_b(5912)^0$      &$\ast\ast\ast$       &  $5912.1\pm0.4$       &  $<0.66$             & $\Xi_b(6095)^-$       &$\cdots$        &  $\cdots$                 &  $\cdots$            & $\cdots$     \\
$\Lambda_b(5920)^0$      &$\ast\ast\ast$       &  $5919.73\pm0.32$     &  $<0.63$             & $\Xi_b(6105)^-$       &$\cdots$        &  $\cdots$                 &  $\cdots$            & $\cdots$     \\
\hline
\end{tabular*}
\end{table*}

For applying the Eqs.~(\ref{eq14}) and~(\ref{eq15}), the diquark
hypothesis will be employed. If the color-magnetic interaction is
the principal reason for diquark correlations, two light quarks in
the heavy baryon systems are expected to correlate strongly. Thus,
they may develop into a diquark because the color-spin interaction
is proportional to the inverse of the quark masses~\cite{Jaffe}.
Furthermore, if the flavor SU(3) symmetry is considered for
\emph{u}, \emph{d}, and \emph{s} quarks,  the total wave functions
of the light diquark should be antisymmetric. Since the spatial and
color parts of diquark are always symmetric and antisymmetric,
respectively, the functions of $|flavor\rangle\times|spin\rangle$
should be symmetric. For this constraint, the scalar diquark ($S =
0$) is always flavor antisymmetric, and the axial-vector ($S = 1$)
flavor symmetric. Following~\cite{Ebert1,Ebert2}, we denote the
scalar diquark as $[q_1, q_2]$, and the axial-vector diquark as
$\{q_1, q_2\}$. Of course, the light quarks in $\Lambda_Q$ and
$\Xi_Q$ baryons are regarded as a scalar diquark.

In the following calculations, $\Lambda_c(2880)$ and $\Xi_c(3080)$
are identified as the 1\emph{D} - wave $\Lambda_c$ and $\Xi_c$
states with the $J^P = 5/2^+$. The $\Xi_c(3055)$ is regarded as the
doublet partner of $\Xi_c(3080)$ with $J^P = 3/2^+$. These three
resonances have been reported by the different
collaborations~\cite{CLEO1,Belle1,Belle2,Belle3,Babar1,Babar2}.
Therefore, the decay properties shall constrain their assignments
strongly. Especially, the ratio of $\Lambda_c(2880)^+$ partial
widths~\cite{Belle2}
\begin{equation}\label{eq16}
\frac{\Gamma(\Lambda_c(2880)\rightarrow\Sigma_c^*(2520)\pi)}{\Gamma(\Lambda_c(2880)\rightarrow\Sigma_c(2455)\pi)}
= (22.5\pm6.2\pm2.5)\%,
\end{equation}
has never been understood well. So it is necessary to discuss their
strong decays for completeness.

\subsection{The EHQ decay formula}
Isgur and Wise noticed that the heavy quark symmetry should play an
important role in the strong decays of heavy-light hadrons a long
time ago~\cite{Wise}. In the $m_Q\rightarrow\infty$ limit, the light
degrees of freedom will decouple from the heavy quark spin. The
transitions between two doublets are governed by a single amplitude
which is proportional to the products of four Clebsch-Gordan
coefficients. The ratios of excited charmed baryons with
negative-parity were predicted by this law~\cite{Wise}. Later, a
more concise formula (the EHQ formula) for the widths of heavy-light
mesons was proposed as
\begin{equation}\label{eq17}
\Gamma^{H\rightarrow H^\prime M}_{j_h,\ell} = (\mathcal
{C}^{s_Q,j'_q,J'}_{j_h,J,j_q})^2\mathcal
{F}^{j_q,j'_q}_{j_h,\ell}(0)p^{2\ell+1}\exp(-\frac{p^2}{\kappa^2})(\frac{M_\rho^2}{M_\rho^2+p^2})^\ell,
\end{equation}
by Eichten, Hill, and Quigg~\cite{Hill}. The normalized coefficient,
$\mathcal {C}^{s_Q,j'_q,J'}_{j_h,J,j_q}$, is denoted by the Wigner
$6j$-symbol,
\begin{eqnarray*}
\mathcal {C}^{s_Q,j'_q,J'}_{j_h,J,j_q}=\sqrt{(2J'+1)(2j_q+1)}\left\{
           \begin{array}{ccc}
                    s_Q  & j'_q & J'\\
                    j_h  & J    & j_q\\
                    \end{array}
     \right\}.
\end{eqnarray*}
Here, \emph{J} and $J'$ represent the total angular momentum of
\emph{H} and $H'$ states. The total angular momentum of the light
degrees of freedom of \emph{H} and $H'$ are $j_q$ and $j_q'$,
respectively. The spins of the heavy quark $Q$ and the light meson
$h$ are denoted by $s_Q$ and $s_h$, then $\vec{j}_h= \vec{s}_h +
\vec{\ell}$. Here, $\ell$ is the orbital angular momentum between
$H'$ and \emph{h}. In their work, the following important
assumptions were included.
\begin{enumerate}[(1)]
 \item The heavy quark symmetry is incorporated by
the Wigner $6j$-symbols. The main breaking of heavy quark symmetry
is expected to be absorbed by the kinematic factors with different
momentums of final states.
 \item The decays of two hadrons in one
doublet are governed by the same transition strengths, $\mathcal
{F}^{j_l,j'_l}_{j_h,\ell}(0)$.
 \item  The $D$, $D_s$, $B$, $B_s$, even \emph{K} mesons are
assumed to be suitable for the EHQ formula.
\end{enumerate}

Then the decays of the $1P (\frac{3}{2}^+)$ and $1D (\frac{5}{2}^-)$
doublets of $D$, $D_s$, $B$ and $B_s$ were studied~\cite{Hill}.
Here, the notation of $nL (j_q^P)$ was used to characterize the
doublet, where \emph{n} is the radial quantum number, \emph{L} is
the orbital angular momentum, and $j_q^P$ refers to the total
angular momentum of the light degrees of freedom and the parity.

One notice that the transition strengths, $\mathcal
{F}^{j_l,j'_l}_{j_h,\ell}(0)$, for the $1P(\frac{3}{2}^+)$ and
$1D(\frac{5}{2}^-)$ doublets do not include the nodal form factors
in the pseudoscalar emission model~\cite{Godfrey}, the $^3P_0$
model~\cite{Kokoski}, and the chiral quark model~\cite{zhong}. In
other words, the node effects may not affect the decays of these
states. However, the node factors can appear in the decays of other
states. For example, a nodal Gaussian factor given by the $^3P_0$
model has been applied for studying whether $D(2637)$ claimed by the
DELPHI Collaboration was a $2S$ state~\cite{Page}.

Recently, we took the EHQ formula to calculate the widths of
2\emph{S}, 1\emph{D} \emph{D} and $D_s$ mesons~\cite{chen1}. The EHQ
formula were also taken to study the \emph{P}-wave heavy-light
mesons~\cite{chen2}. Both decay widths and branching fraction ratios
can be understood well. There the transition strengths were also
extracted by the $^3P_0$ model. The final factor of the
Eq.~(\ref{eq17}) has been omitted in Refs.~\cite{Page,chen1,chen2}
because the momenta of final states are always not large.

For studying the decay of an excited heavy baryons \emph{B} into
another heavy baryon $B'$ and a light meson \emph{M}, we extend the
EHQ formula as
\begin{equation}\label{eq18}
\Gamma^{B\rightarrow B'M} = \xi(\mathcal
{C}^{s_Q,j'_l,J'}_{j_M,j_l,J})^2\mathcal
{F}^{j_l,j'_l}_{j_M,\ell}(0)p^{2\ell+1}\exp(-\frac{5p^2}{8\tilde{\beta}^2}).
\end{equation}
The part of exponential function is obtained by the $^3P_0$ model.
The flavor coefficient, $\xi$, is determined by~\cite{Silvestre}
\begin{eqnarray*}
\xi = \frac{(2I_{B'}+1)(2I_M+1)}{2}\left\{
           \begin{array}{ccc}
                    t_\alpha  & t_\beta & I_B\\
                    I_M       & I_{B'}     & 1/2\\
                    \end{array}
     \right\}^2.
\end{eqnarray*}
$I_B$, $I_{B'}$, $I_M$ are the flavors of \emph{B}, $B'$ and
\emph{M} hadrons, respectively.

The 1\emph{D} $\Lambda_c$ and $\Xi_c$ are governed by two
independent transition strength, $\mathcal {F}^{2,1}_{1,1}(0)$ and
$\mathcal {F}^{2,1}_{3,3}(0)$. In the $^3P_0$ model, they are
written as
\begin{equation}\label{eq19}
\mathcal {F}^{2,1}_{1,1}(0) = \frac{3}{13^2}\mathcal
{G}\frac{1}{\tilde{\beta}^3}(1-\frac{13}{60}\frac{p^2}{\tilde{\beta}^2}),
\end{equation}
and
\begin{equation}\label{eq20}
\mathcal {F}^{2,1}_{3,3}(0) = \frac{1}{2^5\times5^2}\mathcal
{G}\frac{1}{\tilde{\beta}^7}.
\end{equation}
The constant $\mathcal {G}$ is defined as: $\mathcal {G} =
\gamma^2\frac{\tilde{M}_B\tilde{M}_C}{\tilde{M}_A}$, which absorbs
the dimensionless parameter $\gamma$ of the $^3P_0$ model.
$\tilde{M}_A$, $\tilde{M}_B$, and $\tilde{M}_C$ represent the
effective masses for the initial and final
hadrons~\cite{Godfrey,Kokoski}. Obviously, $\mathcal
{F}^{2,1}_{1,1}(0)$ contains a nodal form factor. The node factor
may be important for some particular decay processes. For example,
this kind of node effect has been considered for explaining the
relative branching fractions of $\psi(4040)$~\cite{Yaouanc,Barnes}.
The $\eta_c(4S)$ was predicted to be very narrow due to the node
effect~\cite{He}.

The parameter, $\tilde{\beta}$, has been taken as 0.38 GeV for
studying the \emph{D} and $D_s$ mesons~\cite{chen1,chen2}. For
baryons, the value of $\tilde{\beta}$ should be different. In this
work, we will set it as a variable to investigate whether the
existing decay properties of 1D $\Lambda_c$ and $\Xi_c$ can be
reproduced simultaneously or not.

\subsection{$\Lambda_c^+$ and $\Xi_c^{0,+}$ baryons}
In Eqs.~(\ref{eq14}) and~(\ref{eq15}), there are 8 parameters which
should be fixed, \emph{i.e}., $m_Q$, $m_d$, $\sigma$, $\lambda$,
$v_1$, $v_2$, $\alpha_s$, and $\zeta_Q$. Firstly, we fix the $m_c =
1.470$ GeV, $\sigma_{\Lambda_c} = 1.295$ GeV$^2$, and $m_{[u, d]} +
\zeta_Q = 0.815$ GeV with the spin-averaged masses of $1S$, $1P$,
and $1D$ $\Lambda_c^+$ states. Comparing the value of $m_c$ with the
current-quark mass, 1.275 $\pm$ 0.025 GeV~\cite{PDG}, the velocity
of \emph{c} quark is estimated to be $0.50~\pm~0.03~c$. If we set
$v_2~=~0.5c$, $\zeta_Q = 364$ MeV and $m_{[u,d]} = 451$ MeV are
obtained\footnote{Obviously, it is an approximate method to estimate
the velocity of \emph{c} quark. However, the predicted masses in
Table \ref{tableII} and \ref{tableV} are weekly dependent on the the
velocity. When we vary $v_2$ from $0.40~c$ to $0.62~c$, the
predictions will change no more than $0.1\%$.}. Since the predicted
mass of $1/2^+(2S)$ state are around 2770 MeV (see Table
\ref{tableII}), the $\Lambda_c(2760)^+$ seems to be a candidate of
the first radical excited state of $\Lambda_c(2286)^+$. Thus, the
coefficient $\lambda$ is extracted as 1.57 when $\Lambda_c(2760)^+$
is taken as the $1/2^+(2S)$ state. The light diquark is
ultrarelativistic, which means $v_1\approx1c$. Finally, we fix the
coupling constant, $\alpha$, as 0.67 according with the hyperfine
splitting. The value of $\alpha$ here is consistent with the
QCD-inspired potential model~\cite{Godfrey}. For the $\Xi_c$
baryons, the mass of \emph{c} quark and $\zeta_Q$ fixed above are
taken as the inputs. With the spin-averaged masses of $1S$, $1P$,
and $1D$ states, we fix the $\sigma_{\Xi_c} = 1.558$ GeV$^2$ and
$m_{[u,s]} = 633$ MeV. The masses of scalar diquarks in
Ref.~\cite{Ebert2} were taken as $m_{[u,d]} = 710$ MeV, $m_{[u,s]} =
948$ MeV, which are much larger than our results.

With these parameters in hand, the masses of the $\Lambda_c^+$ and
$\Xi_c$ baryons are shown in Table \ref{tableII} and \ref{tableV}.
The results from other groups~\cite{Ebert2,Roberts2,Capstick} are
also listed for comparison. In Ref.~\cite{Ebert2}, a QCD-motivated
quark potential model was employed and the diquark picture has been
considered. In Ref.~\cite{Roberts2}, the mass spectra were explored
in the three-body picture by the nonrelativistic quark model. Masses
were also studied in the three-body picture by a relativized version
of the quark potential model~\cite{Capstick}. The kinetic term of
quarks and the spin-orbit piece are different in
Refs.~\cite{Roberts2,Capstick}. In Tables \ref{tableII} and
\ref{tableV}, the predicted values in the square brackets belong to
the corresponding states with the $J^P$ listed in the first column.
But the \emph{nL} listed in the parentheses are not the quantum
numbers of these states. We list them for showing the difference
between the two- and three-body pictures of baryons (for details see
Section \ref{sec4}).

\begin{table}
\centering \caption{\label{tableII}The predicted masses of the
$\Lambda_c^+$ baryons (in MeV). We also collect the experimental
values~\cite{PDG} and other theoretical
results~\cite{Ebert2,Roberts2,Capstick} for comparison.}
\label{tableII}
\renewcommand\arraystretch{1.1}
\begin{tabular*}{\columnwidth}{@{\extracolsep{\fill}}cccccc@{}}
\hline
$J^P(nL)$& Exp.~\cite{PDG} &   This work  & Ref.~\cite{Ebert2} & Ref.~\cite{Roberts2} &  Ref.~\cite{Capstick}   \\
\hline
$\frac{1}{2}^+(1S)$& 2286.86     & 2286     & 2286    & 2286    & 2265  \\
$\frac{1}{2}^+(2S)$& 2766.6      & 2766     & 2769    & 2791    & 2775  \\
$\frac{1}{2}^+(3S)$&             & 3112     & 3130    & 3154    & 3170  \\
$\frac{1}{2}^+(4S)$&             & 3397     & 3437    &         &    \\
$\frac{1}{2}^-(1P)$& 2592.3      & 2591     & 2598    & 2625    & 2630   \\
$\frac{3}{2}^-(1P)$& 2628.1      & 2629     & 2627    & 2636    & 2640  \\
$\frac{1}{2}^-(2P)$& \multirow{2}{*}{2939.3}      & 2989     & 2983    &         & [2780]  \\
$\frac{3}{2}^-(2P)$&             & 3000     & 3005    &         & [2840]\\
$\frac{1}{2}^-(3P)$&             & 3296     & 3303    &         & [2830]\\
$\frac{3}{2}^-(3P)$&             & 3301     & 3322    &         & [2885] \\
$\frac{3}{2}^+(1D)$&             & 2857     & 2874    & 2887    & 2910   \\
$\frac{5}{2}^+(1D)$& 2881.53     & 2879     & 2880    & 2887    & 2910    \\
$\frac{3}{2}^+(2D)$&             & 3188     & 3189    & 3120    & 3035   \\
$\frac{5}{2}^+(2D)$&             & 3198     & 3209    & 3125    & 3140    \\
$\frac{5}{2}^-(1F)$&             & 3075     & 3097    & [2872]  & [2900] \\
$\frac{7}{2}^-(1F)$&             & 3092     & 3078    &         & 3125   \\
$\frac{7}{2}^+(1G)$&             & 3267     & 3270    &         & 3175    \\
$\frac{9}{2}^+(1G)$&             & 3280     & 3284    &         &       \\
\hline
\end{tabular*}
\end{table}

Our predictions totally coincide with these results presented by
Ref.~\cite{Ebert2} (see Tables \ref{tableII} and \ref{tableV}). For
these low-lying excited states, both the flux tube model and the
quark potential models can reproduce the masses. The
$\Lambda_c(2595)^+$, $\Lambda_c(2625)^+$, $\Xi_c(2790)^+$, and
$\Xi_c(2815)^+$ are the natural candidates for the 1\emph{P} wave
$\Lambda_c$ and $\Xi_c$ baryons. These assignments were also
supported by the strong decay
analysis~\cite{zhong1,Cheng1,Chen,zhong2}.

As shown in Table \ref{tableII}, the $\Lambda_c(2880)^+$ is a good
candidate of 1\emph{D} state with $J^P~=~5/2^+$. This assignment is
supported by the results of Belle~\cite{Belle2}, that the
$\Lambda_c(2880)^+$ favors $J = 5/2$ over $J = 1/2$ and $J = 3/2$.
However, if this assignment is true, it seems difficult to explain
the ratio of~(\ref{eq16}). When only the contribution of the
\emph{F}-wave partial width was considered , the ratio can be
understood well~\cite{Cheng1}. But the decay channel of
$\Lambda_c(2880)\rightarrow\Sigma_c^*(2520)\pi$ can proceed via
\emph{P}-wave with the lager phase space. In following, we try to
understand the ratio of branching fractions for $\Lambda_c(2880)^+$
by Eq.~(\ref{eq18}).

The possible 1\emph{D} $\Xi_c$ partners of $\Lambda_c(2880)$ are
$\Xi_c(3055)$ and $\Xi_c(3088)$ (see Table \ref{tableV}). All
allowed decay processes of $\Lambda_c(2880)$, $\Xi_c(3055)$, and
$\Xi_c(3088)$ are listed in Table \ref{tableIII}. We also list the
c.m. momentums of the final states and the square of the coefficient
$\mathcal {C}^{s_Q,j'_l,J'}_{j_M,j_l,J}$ in Table \ref{tableIII}.

\begin{table}
\centering \caption{The strong decays of 1\emph{D} $\Lambda_c^+$ and
$\Xi_c^+$. The momentums of final states, \emph{p}, are shown in the
column 4. The values of $(\mathcal {C}^{s_Q,j'_l,J'}_{j_M,j_l,J})^2$
corresponding to \emph{P}- and \emph{F}-wave decays are listed in
the column 5 and 6, respectively. The mass of the predicted state,
$\Lambda_c(2860)^+$, is taken as 2857 MeV. The forbidden decay modes
are marked by ``$\times$''} \label{tableIII}
\renewcommand\arraystretch{1.2}
\begin{tabular*}{\columnwidth}{@{\extracolsep{\fill}}c|ccccc@{}}
\hline
$J^P$ &  Candidates & Decay channels & \emph{p} (MeV)   &  $l = 1$ &  $l = 3$  \\
\hline

\multirow{6}{*}{$\frac{3}{2}^+$}&\multirow{2}{*}{$^\dag\Lambda_c(2860)^+$}&$\Sigma_c(2455)
+ \pi$& 353 & $\frac{5}{6}$    & $\times $  \\
             &     &  $\Sigma_c(2520) + \pi$ & 292 &  $\frac{1}{6}$   & 1   \\
\cline{2-6}
 & \multirow{4}{*}{$\Xi_c(3055)^+$}&$\Sigma_c(2455) + K$&301 & \multirow{2}{*}{$\frac{5}{6}$}    & \multirow{2}{*}{$\times $}  \\
 &    &  $\Xi'_c(2580) + \pi$  &  421  &     &  \\
             &     &  $\Sigma_c(2520) + K$ & 181 &  \multirow{2}{*}{$\frac{1}{6}$}   & \multirow{2}{*}{1}   \\
             &     &  $\Xi_c(2645) + \pi$ & 359 &     &   \\
\hline

\multirow{6}{*}{$\frac{5}{2}^+$}&\multirow{2}{*}{$\Lambda_c(2880)^+$}&$\Sigma_c(2455)
+ \pi$&375 &  $\times $   & $\frac{5}{9}$  \\
             &     &  $\Sigma_c(2520)+\pi$ & 316 &   1  & $\frac{4}{9}$  \\
\cline{2-6}
& \multirow{4}{*}{$\Xi_c(3080)^+$}& $\Sigma_c(2455) + K$&342 &  \multirow{2}{*}{$\times$}  &  \multirow{2}{*}{$\frac{5}{9}$}   \\
             &     &  $\Xi'_c(2580) + \pi$ & 444 &     &    \\
             &     &  $\Sigma_c(2520) + K$ & 236 &   \multirow{2}{*}{1}  & \multirow{2}{*}{$\frac{4}{9}$}   \\
             &     &  $\Xi_c(2645) + \pi$  & 383  &    &    \\
\hline
\end{tabular*}
\end{table}

With the optimal values of $\gamma = 0.45$ and $\tilde{\beta} =$
0.20 GeV, most of the decay properties of 1\emph{D} $\Lambda_c$ and
$\Xi_c$ baryons can be reproduced\footnote{For comparison, we also
listed the results in Table \ref{tableIV} with different $\gamma$
and $\tilde{\beta}$. In the parentheses, the left one corresponds to
$\gamma = 0.30$ and $\tilde{\beta} =$ 0.30 GeV, the right one to
$\gamma = 0.60$ and $\tilde{\beta} =$ 0.25 GeV .} (the second column
of Table \ref{tableIV}). The predicted ratio of
$\Gamma(\Lambda_c(2880)\rightarrow\Sigma_c^*\pi)/\Gamma(\Lambda_c(2880)\rightarrow\Sigma_c\pi)$
is still a little larger than the experimental value, although the
\emph{P}-wave decay is suppressed clearly. At present, only the
Belle Collaboration has measured this branching ratio. So it is
necessary to be confirmed by other Collaborations in future. Another
possible explanation is that the ratio of $\Lambda_c(2880)^+$
reported by Belle may include additional contributions from the
states in the near mass range. In fact, a possible excited
$\Sigma_c$ state ($\Sigma_c(2840)^0$) with a broad structure was
observed by BaBar in 2008~\cite{Babar5}. Its mass and width were
measured to be $2846\pm8\pm10$ MeV and $86^{+33}_{-22}$ MeV,
respectively. The $J^P$ of this state has not been pinned down, but
there was weak evidence that $\Sigma_c(2840)^0$ has $J =
1/2$~\cite{Babar5}. The first radial excitation of $\Sigma_c(2455)$
was predicted in this mass
range~\cite{Ebert1,Ebert2,Capstick,Vijande}. If the
$\Sigma_c(2840)^0$ is a $2S(1/2^+)~\Sigma_c$ state, it can also
decay into $\Sigma_c^*\pi$ and $\Sigma_c\pi$.

\begin{table}
\centering \caption{The predicted decay widths and the ratio of
$\Gamma(\Lambda_c(2880)\rightarrow\Sigma_c^*\pi)/\Gamma(\Lambda_c(2880)\rightarrow\Sigma_c\pi)$.
The experimental values are listed in the column 3 for comparison.
The widths of $\Lambda_c(2880)$, $\Xi_c(3055)$, and $\Xi_c(3080)$
are in units of MeV.}\label{tableIV}
\renewcommand\arraystretch{1.4}
\begin{tabular*}{\columnwidth}{@{\extracolsep{\fill}}lllr@{}}
\hline
  Properties &  $\hspace {0.1cm}$ Prediction & $\hspace {0.1cm}$ Experiment &  $\hspace {0.1cm}$ References  \\
\hline
$\Gamma(\Lambda_c(2880))$                                                                                  & \hspace {0.1cm} 4.5 (3.3/6.5)    &\hspace {0.1cm} 4.7 $\sim$ 6.9    & PDG ~\cite{PDG}     \\
$\frac{\Gamma(\Lambda_c(2880)\rightarrow\Sigma_c^*\pi)}{\Gamma(\Lambda_c(2880)\rightarrow\Sigma_c\pi)}$& \hspace {0.1cm} 1.53 (19.7/3.33) &\hspace {0.1cm} 0.14$\sim$0.31         & Belle ~\cite{Belle2}     \\
$\Gamma(\Xi_c(3055))$                                                                                      & \hspace {0.1cm} 6.0 (2.0/8.4)    &\hspace {0.1cm} 2.0 $\sim$ 16.4    & Belle ~\cite{Belle3}     \\
$\Gamma(\Xi_c(3080))$                                                                                      & \hspace {0.1cm} 8.3 (2.1/11.3)   &\hspace {0.1cm} 0.6 $\sim$ 5.8      & Belle ~\cite{Belle3}     \\
\hline
\end{tabular*}
\end{table}

The $\Xi_c(3055)^+$ signal has been observed in the intermediate
resonant mode of $\Sigma_c(2455)^{++} K^-$, however, no signal in
the $\Sigma_c(2520)^{++} K^-$ decay mode~\cite{Babar1,Belle3}. The
partial width of $\Gamma(\Xi_c(3055)\rightarrow\Sigma_c^*(2520)K)$
is obtained about 0.57 MeV, which seems too small to be observable.
The decay width of the unknown resonance, $\Lambda_c(2860)^+$, is
predicted as 2.8 MeV. The largest decay channel, $\Sigma_c(2520)
\pi$, is about 1.7 MeV. The ratio of
$\Gamma(\Lambda_c(2860)\rightarrow\Sigma_c\pi)/\Gamma(\Lambda_c(2860)\rightarrow\Sigma^*_c\pi)$
is predicted as 0.63. The ratio of branching fraction,
$\Gamma(\Xi_c(3080)\rightarrow\Sigma_c^*
K)/\Gamma(\Xi_c(3080)\rightarrow\Sigma_c K)$, has not been pined
down by experiments~\cite{Babar2}. The theoretical ratio is
predicted as
\begin{equation}
\frac{\Gamma(\Xi_c(3080)\rightarrow\Sigma_c(2455)
K)}{\Gamma(\Xi_c(3080)\rightarrow\Sigma_c^*(2520) K)} \simeq
51.5\%,\nonumber
\end{equation}
which can be test in future. Our calculations indicate that the
channels of $\Xi_c'(2580)\pi$ and $\Xi_c(2645)\pi$ are also
suppressed by the node effect for $\Xi_c(3055)$ and $\Xi_c(3080)$.

\begin{table}
\centering \caption{The predicted masses of the $\Xi_c$ baryons (in
MeV). We also collect the experimental values~\cite{PDG} and other
theoretical results~\cite{Ebert2,Roberts2} for
comparison.}\label{tableV}
\renewcommand\arraystretch{1.1}
\begin{tabular*}{\columnwidth}{@{\extracolsep{\fill}}c|cccc@{}}
\hline
$J^P(nL)$& Exp.~\cite{PDG} &   This work  & Ref.~\cite{Ebert2} & Ref.~\cite{Roberts2}  \\
\hline
$\frac{1}{2}^+(1S)$& 2470.88     & 2467     & 2476    & 2466    \\
$\frac{1}{2}^+(2S)$& 2968.0      & 2959     & 2959    & 2924    \\
$\frac{1}{2}^+(3S)$&             & 3325     & 3323    & [3183]  \\
$\frac{1}{2}^+(4S)$&             & 3629     & 3632    &    \\
$\frac{1}{2}^-(1P)$& 2791.8      & 2779     & 2792    & 2773    \\
$\frac{3}{2}^-(1P)$& 2819.6      & 2814     & 2819    & 2783    \\
$\frac{1}{2}^-(2P)$& \multirow{2}{*}{3122.9}      & 3195     & 3179    &         \\
$\frac{3}{2}^-(2P)$&             & 3204     & 3201    &        \\
$\frac{1}{2}^-(3P)$&             & 3521     & 3500    &         \\
$\frac{3}{2}^-(3P)$&             & 3525     & 3519    &         \\
$\frac{3}{2}^+(1D)$& 3054.2      & 3055     & 3059    & 3012    \\
$\frac{5}{2}^+(1D)$& 3079.9      & 3076     & 3076    & 3004    \\
$\frac{3}{2}^+(2D)$&             & 3407     & 3388    &      \\
$\frac{5}{2}^+(2D)$&             & 3416     & 3407    &      \\
$\frac{5}{2}^-(1F)$&             & 3286     & 3278    &         \\
$\frac{7}{2}^-(1F)$&             & 3302     & 3292    &         \\
$\frac{7}{2}^+(1G)$&             & 3490     & 3469    &         \\
$\frac{9}{2}^+(1G)$&             & 3503     & 3483    &         \\
\hline
\end{tabular*}
\end{table}

The broad $\Lambda_c(2765)^+$ ($\Gamma \approx$ 50 MeV) was first
reported by the CLEO Collaboration ~\cite{CLEO1}. The possible
signal was also seen by Belle Collaboration~\cite{Belle2}. Recently,
Joo \emph{et al}. reanalyzed the full data collected by Belle. They
found that the $\Lambda_c(2765)^+$ was visible in the
$\Sigma_c(2455) \pi$ channel~\cite{Joo}. In the previous fitting
procedure, we have assumed the $\Lambda_c(2765)^+$ to be the
$1/2^+(2S)$ state. For the $\Xi_c$ partner of $\Lambda_c(2765)^+$,
the predicted mass is 2959 MeV (see Table \ref{tableV}), which
supports the assignment of $\Xi_c(2980)^0$ as the first radical
excited state of $\Xi_c(2468)^0$~\cite{Cheng1}.

$\Lambda_c(2940)^+$ and $\Xi_c(3123)^+$ might be the 2\emph{P}
charmed and charm-strange baryons. The predicted masses in our work
and in Ref.~\cite{Ebert2} are about 50$\sim$70 MeV larger than the
experimental values (see Table \ref{tableII} and \ref{tableV}). It
is probably due to the coupled-channel effects. If
$\Lambda_c(2940)^+$ and $\Xi_c(3123)^+$ are 2\emph{P} states, they
can decay through $D^0(D^{*0})p/\Xi_cK$ and $D^+(D^{*+})\Lambda^0$,
respectively, in \emph{S}-wave. Because the $\Xi_cK$ and
$D^{*0,+}P/\Lambda$ thresholds locate nearly below the predicted
values, the coupled-channel effects are expected to be significant.
The coupled-channel effects have been considered as the
responsibility for the anomalously low masses of $D_{s0}^*(2317)$
and $D_{s1}(2460)$~\cite{Close,Hwang,Simonov}.

Recently, Cheng \emph{et al}. examined the invariant-mass spectrum
of $D^0 p$ in $\bar{B}\rightarrow D^0 p \bar{p}$ decays measured by
BaBar~\cite{Babar4}. They found a new charmed baryon resonance with
\begin{equation}
m = 3212 \pm 20 \textrm{MeV};\hspace {1cm} \Gamma = 167 \pm 34
\textrm{MeV},\nonumber
\end{equation}
and denoted it as B$_c(3212)^+$~\cite{Cheng2}. If the B$_c(3212)^+$
is a $\Lambda_c^+$ baryon, it may be a possible 2\emph{D} or
1\emph{F} state according to the predicted masses.

\subsection{$\Lambda_b$ and $\Xi_b$ baryons}
For simplicity, the diquark masses, the coupling constant $\alpha$,
and the values of $\lambda/\zeta_Q$ which have been previously
extracted are taken as inputs for the $\Lambda_b$ and $\Xi_b$
baryons. With the spin-averaged masses of $1S$ and $1P$ $\Lambda_b$
states, we obtained $m_b = 4.804$ GeV  and $\sigma_{\Lambda_b} =
1.147$ GeV$^2$. The mass gap between the spin-averaged masses of
$1P$ $\Lambda_b$ and $\Xi_b$ is also assumed as 182 MeV. Then
$\sigma_{\Xi_b} = 1.386$ GeV$^2$ is found. With these values, the
masses of orbital excited $\Lambda_b^{0}$ and $\Xi_b^{0,-}$ are
predicted, separately, in Table \ref{tableVI} and \ref{tableVII}.

\begin{table}
\centering \caption{The predicted masses of the $\Lambda_b^0$
baryons (in MeV). We also collect the experimental values~\cite{PDG}
and other theoretical results~\cite{Ebert2,Roberts2,Capstick} for
comparison.}\label{tableVI}
\renewcommand\arraystretch{1.1}
\begin{tabular*}{\columnwidth}{@{\extracolsep{\fill}}c|ccccc@{}}
\hline
$J^P(nL)$& Exp.~\cite{PDG} &   This work  & Ref.~\cite{Ebert2} & Ref.~\cite{Roberts2} &  Ref.~\cite{Capstick}  \\
\hline
$\frac{1}{2}^+(1S)$& 5619.4     & 5619     & 5620    & 5612    & 5585    \\
$\frac{1}{2}^-(1P)$& 5912.0     & 5911     & 5930    & 5939    & 5912   \\
$\frac{3}{2}^-(1P)$& 5919.8     & 5920     & 5942    & 5941    & 5920    \\
$\frac{3}{2}^+(1D)$&            & 6147     & 6190    & 6181    & 6145    \\
$\frac{5}{2}^+(1D)$&            & 6153     & 6196    & 6181    & 6165    \\
$\frac{5}{2}^-(1F)$&            & 6346     & 6408    & 6206    & 6205    \\
$\frac{7}{2}^-(1F)$&            & 6351     & 6411    &         & 6360     \\
$\frac{7}{2}^+(1G)$&            & 6523     & 6598    & 6433    & 6445   \\
$\frac{9}{2}^+(1G)$&            & 6526     & 6599    &         & 6580    \\
\hline
\end{tabular*}
\end{table}

Two narrow states, named $\Lambda_b(5912)^0$ and
$\Lambda_b(5920)^0$, were observed in the $\Lambda_b^0\pi^+\pi^-$
spectrum by LHCb~\cite{LHCb1}. The $\Lambda_b(5920)^0$ was later
confirmed by CDF Collaboration~\cite{CDF1}. The hyperfine of
$\Lambda_b(5912)^0$ and $\Lambda_b(5920)^0$ given by equation
\ref{eq15} is consistent with the experiments. The
$\Lambda_b(5912)^0$ and $\Lambda_b(5920)^0$ are the good candidates
for 1\emph{P} $\Lambda_b$ baryons with $J^P = 1/2^-$ and $3/2^-$,
respectively.

\begin{table}
\centering \caption{The predicted masses of the $\Xi_b^0$ baryons
(in MeV). We also collect the experimental values~\cite{PDG} and
other theoretical results~\cite{Ebert2,Roberts2} for
comparison.}\label{tableVII}
\renewcommand\arraystretch{1.1}
\begin{tabular*}{\columnwidth}{@{\extracolsep{\fill}}c|cccc@{}}
\hline
$J^P(nL)$& Exp.~\cite{PDG} &   This work  & Ref.~\cite{Ebert2} & Ref.~\cite{Roberts2}  \\
\hline
$\frac{1}{2}^+(1S)$& 5795.8     & 5801     & 5803    & 5806     \\
$\frac{1}{2}^-(1P)$&            & 6097     & 6120    & 6090     \\
$\frac{3}{2}^-(1P)$&            & 6106     & 6130    & 6093     \\
$\frac{3}{2}^+(1D)$&            & 6344     & 6366    & 6311     \\
$\frac{5}{2}^+(1D)$&            & 6349     & 6373    & 6300     \\
$\frac{5}{2}^-(1F)$&            & 6555     & 6577    &          \\
$\frac{7}{2}^-(1F)$&            & 6559     & 6581    &          \\
$\frac{7}{2}^+(1G)$&            & 6743     & 6760    &          \\
$\frac{9}{2}^+(1G)$&            & 6747     & 6762    &          \\
\hline
\end{tabular*}
\end{table}

The 1\emph{P} and higher excitations of $\Xi_b$ have not been
observed so far. The predictions presented in Table \ref{tableVII}
will be helpful to the future experimental searches. One notice that
the predictions of two 1\emph{P} $\Xi_b$ baryons here and in
Refs.~\cite{Ebert2,Roberts2} are above the $\Xi'_b(5945) \pi$
threshold. Thus, the 1\emph{P} $\Xi_b$ states can be searched in the
channel of $\Xi'_b(5945) \pi$.

\section{Further Discussions}
\label{sec4}
\subsection{Distinctions between the heavy quark-light diquark and three-body pictures for heavy baryons}
In the Section \ref{sec3}, we have studied the mass spectrum and
decay properties of heavy baryons in the heavy quark-light diquark
picture. All $\Lambda_Q$ and $\Xi_Q$ states can be understood.
However, we still cannot exclude the three-body picture for the
heavy baryons. In fact, there exist other possible mechanisms for
the ``missing resonances problem''. As an example, the authors of
Refs.~\cite{Koniuk,Roberts1} pointed out that the missing $N^*$ and
$\Delta$ resonances were due to the weak couplings to the $N \pi$
channel which was used predominantly for production of excited $N^*$
and $\Delta$ baryons. In deed, some experimental evidences against
the quark-diquark for $N^*$ resonances~\cite{TAPS}. So what can be
used as criteria to distinguish these two pictures for heavy
baryons? In this section, we will take the $\Lambda_c$ baryons to
illustrate this issue.

Firstly, the masses of $\Lambda_c$ are predicted to be strikingly
different in the diquark and three-quark models. In the second row
of Table~\ref{tableVIII}, we list the \emph{S-}, \emph{P-},
\emph{D-}, and \emph{F-} states in the diquark model. Besides these
states, there are other possible excitations in three-quark model,
which are listed in the rows 3 to 9. Details of these denotations
can be found in Ref.~\cite{Chen}. In the diquark models, the lowest
excitation with $J^P = 5/2^-$ is a \emph{F}-wave state, which is
denoted as $\Lambda_{c3}(5/2^-)$. Differently, the three-quark
models allow a \emph{P}- wave state to be $5/2^-$. Here, we denote
it as $\tilde{\Lambda}_{c2}(5/2^-)$. The predicted masses of
$\tilde{\Lambda}_{c2}(5/2^-)$ state is in the range of
2870$\sim$2900 MeV~\cite{Roberts2,Capstick}, which is much lower
than the predictions of $\Lambda_{c3}(5/2^-)$ in diquark models (see
Table \ref{tableII}). In fact, the states with negative parity are
not allowed to locate in the energy range from 2650 to 2930 MeV in
the diquark models~\cite{Crede}. If any state of $\Lambda_c$ was
found to have negative parity in this mass range, the hypothesis of
the quark-diquark picture would be excluded.

\begin{table}
\centering \caption{Allowed $\Lambda_c$ states in the quark-diquark
and three-body picture.}\label{tableVIII}
\renewcommand\arraystretch{1.2}
\begin{tabular*}{\columnwidth}{@{\extracolsep{\fill}}lllll@{}}
\hline
\emph{S}  & \emph{P} &   \emph{D}  & \emph{F} & $\cdots$  \\
\hline
$\Lambda_{c0}(\frac{1}{2}^+)$& $\Lambda_{c1}(\frac{1}{2}^-,~\frac{3}{2}^-)$    & $\Lambda_{c2}(\frac{3}{2}^+,~\frac{5}{2}^+)$     & $\Lambda_{c3}(\frac{5}{2}^-,~\frac{7}{2}^-)$    & $\cdots$     \\
\hline
 None & $\tilde{\Lambda}_{c0}(\frac{1}{2}^-)$          & $\hat{\Lambda}_{c2}(\frac{3}{2}^+,~\frac{5}{2}^+)$     & $\cdots$    & $\cdots$     \\
 & $\tilde{\Lambda}_{c1}(\frac{1}{2}^-,~\frac{3}{2}^-)$& $\check{\Lambda}^1_{c0}(\frac{1}{2}^+)$                &          &      \\
 & $\tilde{\Lambda}_{c2}(\frac{3}{2}^-,~\frac{5}{2}^-)$& $\check{\Lambda}^1_{c1}(\frac{1}{2}^+,~\frac{3}{2}^+)$     &      &      \\
 &                                                     & $\check{\Lambda}^1_{c2}(\frac{3}{2}^+,~\frac{5}{2}^+)$     &      &     \\
 &                                                     & $\check{\Lambda}^0_{c1}(\frac{1}{2}^+,~\frac{3}{2}^+)$     &      &         \\
 &                                                     & $\check{\Lambda}^2_{c2}(\frac{3}{2}^+,~\frac{5}{2}^+)$     &      &         \\
 &                                                     & $\check{\Lambda}^2_{c3}(\frac{5}{2}^+,~\frac{7}{2}^+)$     &      &         \\
\hline
\end{tabular*}
\end{table}

Secondly, we point out that the decay properties of the
$\tilde{\Lambda}_{c2}(5/2^-)$ and $\Lambda_{c3}(5/2^-)$ states are
also different because their total angular momentum of the light
degrees of freedom, $j_l$, are different. In the heavy quark limit,
the following two model-independent ratios are shown for the
$\tilde{\Lambda}_{c2}(5/2^-)$ and $\Lambda_{c3}(5/2^-)$ states by
Eq.~(\ref{eq18}).

For three-body picture,
\begin{equation}\label{eq21}
\tilde{R} =
\frac{\Gamma(\tilde{\Lambda}_{c2}(\frac{5}{2}^-)\rightarrow\Sigma_c(2455)\pi)}{\Gamma(\tilde{\Lambda}_{c2}(\frac{5}{2}^-)\rightarrow\Sigma_c(2520)\pi)}
= \frac{2}{7} < 1.
\end{equation}

For diquark picture,
\begin{equation}\label{eq22}
R =
\frac{\Gamma(\Lambda_{c3}(\frac{5}{2}^-)\rightarrow\Sigma_c(2455)\pi)}{\Gamma(\Lambda_{c3}(\frac{5}{2}^-)\rightarrow\Sigma_c(2520)\pi)}
= \frac{7}{2} > 1.
\end{equation}

Here,we ignore the decay channel of $\Sigma_c(2520)\pi$ in
\emph{G}-wave for the small phase space. In practice, the heavy
quark symmetry is broken for the finite mass of $m_Q$. When the
different momentums of the final states are considered, the ratio of
$\tilde{R}$ is predicted about 0.54$\sim$0.59. However, the ratio of
$R$ is about 4.56 which is much large. Here the predicted masses
shown in Table \ref{tableII} have been used.

Finally, we stress that $\tilde{\Lambda}_{c2}(5/2^-)$ is a nice
criterion to test the diquark picture for charmed baryons. (1). The
mass of $\tilde{\Lambda}_{c2}(5/2^-)$ is not very high for the
further experiments. (2). $\tilde{\Lambda}_{c2}(5/2^-)$ is the only
state with $J^P = 5/2^-$ in the range of 2760 to 2900 MeV. So the
mixing effects is insignificant. (3). The primary decay channels of
$\tilde{\Lambda}_{c2}(5/2^-)$ and $\Lambda_{c3}(5/2^-)$ are
$\Sigma_c(2455)\pi$ and $\Sigma_c(2520)\pi$ which are used
predominantly to search for the high excited $\Lambda_c^+$ states.

\subsection{Mass gaps between $\Lambda_Q$ and $\Xi_Q$ baryons}
In this subsection, the mass gaps between $\Lambda_Q$ and $\Xi_Q$
will be explained by the Eqs.~\ref{eq14} and~\ref{eq15}. We take the
$\Lambda_c/\Xi_c$ as example, and show that the main mass gap
between the corresponding states originates from the different
masses of diquark. To this end, we combine these two equations in
the form
\begin{equation}\label{eq23}
\varepsilon = M_Q + m_d + \bar{\Lambda}_{nL} (\kappa n + L) +
V^{so}_{nL}~\hat{s}_Q\cdot\hat{L},
\end{equation}
where
\begin{equation}
\begin{split}
&M_Q = m_Q + \zeta_Q;~~~~\bar{\Lambda}_{nL} =
\frac{\sigma}{2(\varepsilon - m_Q + m_d +\zeta_Q)};~~~~\\&\rm {and},
~~~\emph{V}^{\emph{so}}_{\emph{nL}} =
\frac{\alpha_\emph{s}}{3\times2^{5/2}}~(\frac{\sigma}{\kappa
\emph{n} +
\emph{L}})^{3/2}\frac{1}{\emph{m}_\emph{d}\emph{m}_\emph{Q}}.\nonumber
\end{split}
\end{equation}

We treated $\zeta_c$ and $m_c$ as constants in our calculations.
Then $M_c$ are equal for $\Lambda_c$ and $\Xi_c$. We define
$H_d\equiv\bar{\Lambda}_{nL} (\kappa n + L)$ which could be regarded
as the bound energy of diquark in a heavy baryon system. The values
of $H_d$ are shown for different excited $\Lambda_c$ and $\Xi_c$
states in Table \ref{tableIX}. The hyperfine splittings,
$H^{so}_{nL}$, are also presented for these states, which are caused
by the spin-orbit coupling. Obviously, the differences of
$\bar{\Lambda}_{nL}$ and $V^{so}_{nL}$ are small for $\Lambda_c$ and
$\Xi_c$, which reflect the similar dynamics between $\Lambda_c$ and
$\Xi_c$ families.

\begin{table}
\centering \caption{The values of $H_d$ and $H^{so}_{nL}$ for the
2\emph{S}, 1\emph{P}, 2\emph{P}, 3\emph{P}, 1\emph{D}, 2\emph{D},
1\emph{F}, and 1\emph{G} $\Lambda_c/\Xi_c$ states (in units of
MeV).}\label{tableIX}
\renewcommand\arraystretch{1.2}
\begin{tabular*}{\columnwidth}{@{\extracolsep{\fill}}c|cc|cc@{}}
\hline
\multirow{2}{*}{states (n\emph{L})}&  \multicolumn{2}{c|}{$H_d$}  & \multicolumn{2}{c}{$H^{so}_{nL}$} \\
\cline{2-5}
 & $\Lambda_c$ & $\Xi_c$ &  $\Lambda_c$   &  $\Xi_c$ \\
\hline
2 \emph{S}  & 481   & 492   & $-$      &  $-$   \\
1 \emph{P}  & 330   & 335   & 38       &  35   \\
2 \emph{P}  & 711   & 733   & 11       &  9   \\
3 \emph{P}  & 1014  & 1057  & 5        &  4   \\
1 \emph{D}  & 585   & 601   & 22       & 21   \\
2 \emph{D}  & 910   & 946   & 10       & 9   \\
1 \emph{F}  & 800   & 828   & 17       &  16   \\
1 \emph{G}  & 989   & 1030  & 13       &  13   \\
\hline
\end{tabular*}
\end{table}

It is interesting to compare the Eq.~(\ref{eq23}) with the formula
given by the heavy-quark effective theory (HQET). In HQET, the mass
formula for heavy baryons is written as~\cite{HQET}
\begin{equation}\label{eq24}
\hspace {1.2cm}\varepsilon = m_Q + \frac{\tilde{a}}{m_Q} + m_d + H_d
+ H_{hyp}
\end{equation}
Here, $m_Q$ and $m_d$ are the masses of heavy quark and
light-diquark, respectively. The second term, $\tilde{a}/m_Q$,
arises from the kinetic energy of the heavy quark inside the heavy
baryons. $H_d$ denotes the bound energy of diquark in baryon
systems. $H_{hyp}$ represents the hyperfine interactions. The bound
energy, $H_d$, can not be given in the HQET framework. Comparing the
Eqs.~(\ref{eq23}) and~(\ref{eq24}), we find that $H_d$ is equal to
$\bar{\Lambda}_{nL} (\kappa n + L)$ in the RFT model.

\section{\label{V}Summary and Outlook}
We have analytically derived a mass formula for the excited
heavy-light hadrons within the relativistic flux tube model. Then
the formula is applied to study the spectra of $\Lambda_Q$ and
$\Xi_Q$ (\emph{Q} = \emph{c} or \emph{b} quark) baryons, where the
heavy quark-light diquark picture is considered. The spin-orbit
interaction was borrowed directly from the QCD-motivated potential
models. Our results support the main conclusion of
Ref.~\cite{Ebert2} that the available $\Lambda_Q$ and $\Xi_Q$
baryons can be understood well in the heavy quark-light diquark
picture. But the masses of light diquark obtained in our works are
much smaller. Our results indicate that $\textbf{s}_Q \cdot
\textbf{L}$ is the main spin-dependent term for $\Lambda_Q$ and
$\Xi_Q$ baryons

In the heavy quark-light diquark picture, $\Lambda_c(2760)^+$ and
$\Xi_c(2980)$ can be assigned to the 2\emph{S} sates with $J^P =
1/2^+$. $\Lambda_c(2940)^+$ and $\Xi_c(3123)$ might be the $2P$
excitations of $\Lambda_c$ and $\Xi_c$. The $\Lambda_c(2880)^+$ and
$\Xi_c(3080)$ are the $1D$ $\Lambda_c$ and $\Xi_c$ states with $J^P
= 5/2^+$. The $\Xi_c(3055)$ could be the doublet partner of
$\Xi_c(3080)$ with $J^P = 3/2^+$. We assign the new resonance,
B$_c$(3212), as a 2\emph{D} or 1\emph{F} $\Lambda_c^+$ states
temporarily. The $\Lambda_b(5912)^0$ and $\Lambda_b(5920)^0$ are the
1\emph{P} bottom baryons with $J^P = 1/2^-$ and $3/2^-$,
respectively.

It is important to emphasize that these assignments seem not to
contradict their strong decay properties. The narrow structures of
$\Lambda_c(2880)^+$, $\Xi_c(3055)$, and $\Xi_c(3080)$ have been
understood. The node effects may be significant for the decays of
1\emph{D} $\Lambda_c$ and $\Xi_c$. We partially interpreted the
ratio of $\Lambda_c(2880)^+$ partial widths, which was measured by
Belle. The ratio of $\Gamma(\Xi_c(3080)~\rightarrow~\Sigma_c(2455)
K)/\Gamma(\Xi_c(3080)~\rightarrow~\\\Sigma_c^*(2520) K)$ is
predicted about $51.5\%$, which can be tested in future. Surely, a
systematical study of decays of $\Lambda_c$ and $\Xi_c$ baryons in
the heavy quark-light diquark picture is necessary in further
research.

At present, we still cannot exclude the three-body picture for the
heavy baryons. The distinctions between the heavy quark-light
diquark and three-body pictures were discussed. We propose a search
for the $\tilde{\Lambda}_{c2}(\frac{5}{2}^-)$ state which can help
to distinguish the diquark and three-body schemes. In a word, the
investigations in this work are expected to be helpful for the heavy
baryon physics.

\begin{acknowledgements}
Bing Chen thanks T.J. Burns for the valuable suggestions. This work
is supported by the National Natural Science Foundation of China
under grant Nos. 11305003, 11475111, 11475004, U1404114, and
U1204115. It is also supported by the Key Program of the He'nan
Educational Committee of China under grant Nos.13A140014 and
14-A140016, the Innovation Program of Shanghai Municipal Education
Commission under grant No. 13ZZ066.
\end{acknowledgements}


\begin{thebibliography}{90}

\bibitem{PDG}
K. A. Olive \emph{et al}. (Particle Data Group), Chin. Phys. C,
\textbf{38}, 090001 (2014).

\bibitem{Sant1}
G. Galat$\grave{a}$, and E. Santopinto, Phys. Rev. C \textbf{86},
045202 (2012).

\bibitem{Klempt}
E. Klempt and J.-M. Richard, Rev. Mod. Phys. \textbf{82}, 1095
(2010).

\bibitem{Crede}
V. Crede and W. Roberts, Rep. Prog. Phys. \textbf{76}, 076301
(2013).

\bibitem{Sant2}
E. Santopinto, Phys. Rev. C \textbf{72}, 022201 (2005).

\bibitem{Sant3}
J. Ferretti, A. Vassallo, and E. Santopinto, Phys. Rev. C
\textbf{83}, 065204 (2011).

\bibitem{Gutierreza}
  C.~Gutierrez and M.~De Sanctis,
  Eur.\ Phys.\ J.\ A {\bf 50}, 169 (2014).

\bibitem{Ebert1}
D. Ebert, R. N. Faustov, and V. O. Galkin, Phys. Lett. B
\textbf{659}, 612 (2008).

\bibitem{Ebert2}
D. Ebert, R. N. Faustov, and V. O. Galkin, Phys. Rev. D \textbf{84},
014025 (2011).

\bibitem{Olsson1}
D. LaCourse and M. G. Olsson, Phys. Rev. D \textbf{39}, 2751 (1989).

\bibitem{Olsson2}
C. Olson, M. G. Olsson, and K. Williams, Phys. Rev. D \textbf{45},
4307 (1992).

\bibitem{Selem}
A. Selem and F. Wilczek, arXiv: hep-ph/ 0602128.

\bibitem{zhang1}
H.-Y. Shan and A. Zhang, Chin. Phys. C \textbf{34}, 16 (2010).

\bibitem{zhang2}
B. Chen, D.-X. Wang and A. Zhang, Phys. Rev. D \textbf{80}, 071502
(2009).

\bibitem{zhang3}
B. Chen, D.-X. Wang and A. Zhang, Chin. Phys. C \textbf{33}, 1327
(2009).

\bibitem{Nambu}
Y. Nambu, ``\emph{Quark model and the factorization of the Veneziano
Amplitude},'' in \emph{Symmetries and quark models}, edited by R.
Chand (Gordon and Breach, New York, 1970).

\bibitem{Susskind}
L. Susskind, Nuovo Cim. A \textbf{69}, 457 (1970).

\bibitem{Goto}
T. Goto, Prog. Theor. Phys. \textbf{46}, 1560 (1971) .

\bibitem{Olsson3}
C. Olson, M. G. Olsson, and D. La Course, Phys. Rev. D \textbf{49},
4675 (1994).

\bibitem{Olsson4}
M. G. Olsson and S. Veseli, Phys. Rev. D \textbf{51}, 3578 (1995).

\bibitem{Olsson5}
M. G. Olsson, S. Veseli, and K. Williams, Phys. Rev. D \textbf{53},
4006 (1996).

\bibitem{Olsson6}
T. J. Allen, M. G. Olsson, and S. Veseli, Phys. Rev. D \textbf{60},
074026 (1999).

\bibitem{Olsson7}
T. J. Allen, M. G. Olsson, and J. R. Schmidt, Phys. Rev. D
\textbf{69}, 054013 (2004).

\bibitem{Olsson8}
T. J. Allen, M. G. Olsson, and S. Veseli, Phys. Rev. D \textbf{62},
094021 (2000).

\bibitem{Burns}
T. J. Burns, F. Piccinini, A. D. Polosa, and C. Sabelli, Phys. Rev.
D \textbf{82}, 074003 (2010).

\bibitem{penta}
M. Iwasaki and F. Takagi, Phys. Rev. D \textbf{77}, 054020 (2008).

\bibitem{glue}
F. Buisseret, V. Mathieu, and C. Semay, Phys. Rev. D \textbf{80},
074021 (2009).

\bibitem{Brambilla}
N. Brambilla, G. M. Prosperi, and A. Vairo, Phys. Lett. B
\textbf{362}, 113 (1995).

\bibitem{Baker}
M. Baker and R. Steinke, Phys. Rev. D \textbf{63},  094013 (2001).

\bibitem{Cahn}
R. N. Cahn and J. D. Jackson, Phys. Rev. D \textbf{68}, 037502
(2003).

\bibitem{Jaffe}
R. L. Jaffe, Phys. Rep. \textbf{409}, 1 (2005).

\bibitem{CLEO1}
M. Artuso \emph{et al}. (CLEO Collaboration), Phys. Rev. Lett.
\textbf{86}, 4479 (2001).

\bibitem{Belle1}
R. Chistov \emph{et al}. (Belle Collaboration), Phys. Rev. Lett.
\textbf{97}, 162001, (2006).

\bibitem{Belle2}
K. Abe \emph{et al}. (Belle Collaboration), Phys. Rev. Lett.
\textbf{98}, 262001, (2007).

\bibitem{Belle3}
Y. Kato \emph{et al}. (Belle Collaboration), Phys. Rev. D
\textbf{89}, 052003 (2014).

\bibitem{Babar1}
B. Aubert \emph{et al}. (BaBar Collaboration), Phys. Rev. Lett.
\textbf{98}, 012001, (2007).

\bibitem{Babar2}
B. Aubert \emph{et al}. (BaBar Collaboration), Phys. Rev. D
\textbf{77},012002 (2008).

\bibitem{Wise}
N. Isgur and M. B. Wise, Phys. Rev. Lett. \textbf{66}, 1130 (1991).

\bibitem{Hill}
E. J. Eichten, C. T. Hill, and C. Quigg, Phys. Rev. Lett.
\textbf{71}, 4116 (1993).

\bibitem{Godfrey}
S. Godfrey and N. Isgur, Phys. Rev. D \textbf{32}, 189 (1985).

\bibitem{Kokoski}
R. Kokoski and N. Isgur, Phys. Rev. D \textbf{35}, 907 (1987).

\bibitem{zhong}
X.-H. Zhong and Q. Zhao, Phys. Rev. D \textbf{78}, 014029 (2008).

\bibitem{Page}
P. R. Page, Phys. Rev. D \textbf{60}, 057501 (1999).

\bibitem{chen1}
B. Chen, L. Yuan, and A. Zhang, Phys. Rev. D \textbf{83}, 114025
(2011).

\bibitem{chen2}
B. Chen, L. Yuan, and A. Zhang, arXiv:1210.6151.

\bibitem{Silvestre}
B. Silvestre-Brac, Phys. Rev. D \textbf{46}, 2179 (1992).

\bibitem{Yaouanc}
A. Le Yaouanc, L. Oliver, O. P$\grave{e}$ne, and J.-C. Raynal, Phys.
Lett. B \textbf{71}, 397 (1977).

\bibitem{Barnes}
T. Barnes, S. Godfrey, and E. S. Swanson, Phys. Rev. D \textbf{72},
054026 (2005).

\bibitem{He}
L.-P. He, D. Y. Chen, X. Liu and T. Matsuki, Eur. Phys. J. C
\textbf{74}, 3208 (2014).

\bibitem{Roberts2}
W. Roberts and M. Pervin, Int. J. Mod. Phys. A \textbf{23}, 2817
(2008).

\bibitem{Capstick}
S. Capstick and N. Isgur, Phys. Rev. D \textbf{34}, 2809 (1986).

\bibitem{zhong1}
X.-H. Zhong and Q. Zhao, Phys. Rev. D \textbf{77}, 074008 (2008).

\bibitem{zhong2}
L.-H. Liu, L.-Y. Xiao, and X.-H. Zhong, Phys. Rev. D \textbf{86},
034024 (2012).

\bibitem{Chen}
C. Chen, X.-L. Chen, X. Liu, W.-Z. Deng, and S.-L. Zhu, Phys. Rev. D
\textbf{75}, 094017 (2007).

\bibitem{Cheng1}
H. Y. Cheng and C. K. Chua, Phys. Rev. D \textbf{75}, 014006 (2007).

\bibitem{Babar5}
B. Aubert \emph{et al}. (BaBar Collaboration), Phys. Rev. D
\textbf{78},112003 (2008).

\bibitem{Vijande}
A. Valcarce, H. Garcilazo, and J. Vijande, Eur. Phys. J. A
\textbf{37}, 217 (2008).

\bibitem{Joo}
C. W. Joo, Y. Kato, and K. Tanida, and Y. Kato, PoS Hadron
\textbf{201}, 2013 (2014).

\bibitem{Close}
T. Barnes, F. E. Close, and H. J. Lipkin, Phys. Rev. D \textbf{68},
054006 (2003).

\bibitem{Hwang}
D. S. Hwang and D. W. Kim, Phys. Lett. B \textbf{601}, 137 (2004).

\bibitem{Simonov}
Yu.A. Simonov and J. A. Tjon, Phys. Rev. D \textbf{70}, 114013
(2004).

\bibitem{Babar4}
P. del Amo Sanchez \emph{et al}. (BaBar Collaboration), Phys. Rev. D
\textbf{85}, 092017 (2012).

\bibitem{Cheng2}
H. Y. Cheng, C. Q. Geng, and Y. K. Hsiao, Phys. Rev. D \textbf{89},
034005 (2014).

\bibitem{LHCb1}
R. Aaij \emph{et al}. (LHCb Collaboration), Phys. Rev.
Lett.\textbf{109}, 172003 (2012).

\bibitem{CDF1}
T. Aaltonen \emph{et al}. (CDF Collaboration), Phys. Rev.
Lett.\textbf{88}, 071101 (2013).

\bibitem{Koniuk}
R. Koniuk and N. Isgur, Phys. Rev. D \textbf{21}, 1868 (1980); Phys.
Rev. D \textbf{23}, 818(E) (1981).

\bibitem{Roberts1}
S. Capstick and W. Roberts, Phys. Rev. D \textbf{47}, 1994 (1993).

\bibitem{TAPS}
A. Thiel \emph{et al}. (CBELSA/TAPS Collaboration), Phys. Rev. Lett.
\textbf{114}, 091803 (2015).

\bibitem{HQET}
Fayyazuddin and Riazuddin, \emph{A Modern Introduction to Particle
Physics}, (World Scientific Publishing Co. Pte. Ltd., Singapore,
2000, 2nd edition), p.717.


\end{thebibliography}
\end{document}